# Observation of Superfluorescence from a Spontaneous Coherence of Excitons in ZnTe Crystal: Evidence for Bose-Einstein Condensation of Excitons?

D. C. Dai and A. P. Monkman
Department of Physics, Durham University, South Road, Durham DH1 3LE, United Kingdom
e-mails: dechang.dai@durham.ac.uk and a.p.monkman@durham.ac.uk

Superfluorescence (SF) is the emission from a dense coherent system in population inversion, formed from an initially incoherent ensemble. This is characterised by an induction time ($\tau_D$) for the spontaneous development of the macroscopic quantum coherence. Here we report detailed observation of SF on ultrafast timescale from a quantum ensemble of coherent excitons in highly excited intrinsic bulk ZnTe single crystal at 5 K, showing a characteristic $\tau_D$ from 40 ps to 10 ps, quantum noise and fluctuations, and quantum beating and ringing. From this clear observation of SF from a spontaneous coherence of excitons we infer that this is indicative of the formation of BEC of excitons on an ultrafast timescale.

Superfluorescence (SF) and superradiance (SR) are both forms of *cooperative emission* arising from a dense *coherent* ensemble ($N_c$) in a population inversion ($N$), first predicted by Dicke in 1954 [1], and subsequently termed by Bonifacio and Lugiato in 1975 [2]. If a dense population inversion has an initial macroscopic polarization, as created coherently by a laser ($N=N_c$) for example, the resultant emission is SR. However, in some cases, an initially *incoherent* dense $N$ can form *spontaneous coherence* over $N_c$ with a unique characteristic induction time, $\tau_D$, resulting in SF [2,3]. In contrast, amplified spontaneous emission (ASE) is the *collective emission* from a purely *incoherent* dense $N$ [4]. Although SF, SR and ASE are all observed as mirrorless "lasing" in the absence of a laser cavity, SF differs essentially from SR and ASE by the existence of $\tau_D$, for the development of macroscopic spontaneous quantum coherence [2-4]. In order to resolve the SF process in experiment, both the duration of excitation pulse ($\tau_p$) and the time resolution of emission detection system ($\Delta t$) must be shorter than $\tau_D$, i.e. ($\tau_p$, $\Delta t$)<$\tau_D$ [5-11]. Also, SF has the observable features of emission line narrowing, greatly increased intensity, quantum noise and quantum fluctuations, quantum beating and ringing *etc*. It also follows the relationships: $I_{SF} \propto N_c^2$, $\tau_R=(8\pi/3N_c\lambda^2 l)\tau_{SP}$, and $\tau_D=\tau_R[\ln(2\pi S/N_c)^{1/2}]^2/4$, where $\tau_{SP}$ is the lifetime of spontaneous emission, $I_{SF}$ and $\tau_R$ are respectively the intensity and characteristic radiation time of SF emission at the wavelength $\lambda$, $S$ and $l$ are the area and thickness of the gain medium respectively [1-10].

SF emission from the whole coherent ensemble of $N_c$ is triggered by a random individual spontaneous emission event within $N_c$ due to quantum fluctuations [6-9], therefore the properties of SF (such as $\tau_D$, pulse shape and intensity) fluctuate from shot to shot, i.e. exhibiting quantum noise [3,5,6]. SF is particularly interesting because it is intrinsically a quantum mechanical phenomenon and provides a significant tool to study the quantum coherence and the macroscopic quantum fluctuations in the time domain [3]. Importantly, $\tau_D$ is always limited by the dephasing time $T_2$ or $T_2^*$, which always act to destroy the coherence [6-10], and are defined by the inverse of the transition cross-sections in homogeneous or inhomogeneous systems respectively. Thus SF has a key criterion: $(\tau_R\tau_D)^{1/2}<T_2^*$ [7-9].

Given that SF is the signature of a spontaneous development of a coherent exciton ensemble, we think this can be viewed as a Bose-Einstein condensation (BEC) of the excitons through spontaneous symmetry breaking. The convincing observation of an excitonic BEC in a semiconductor is a long held but unresolved issue from its prediction in 1962 [12-16]. Practically all of the previous claims have been disputed due to the lack of unique evidence [15,16]. One of the key issues is the experimental proof of the formation of the spontaneous macroscopic Bose coherence [16], which has been well attested in superfluidity, superconductivity and atomic BEC. A drawback in experiments is the short lifetime of the excitons, which is on ultrafast timescale (usually on the order of magnitude of 100 ps), within which the condensation is expected to occur [12-15]. SF from a quantum coherent ensemble of excitons should be an excellent way of demonstrating the BEC of excitons, however, this has not been previously suggested in theoretical work nor has it been successfully observed in semiconductor so far. Here we directly address this issue.

From the first observation of SF from highly excited HF gas [6], many other gaseous atomic or molecular systems, with predominantly homogeneous line-broadening, have been unambiguously found to emit SF [5-9]. There are only a few reports of SF in the crystalline solid phase with inhomogeneous broadening, of which the most documented is KCl:$O_2^-$ [10,11], Jho *et al* reported observations of SF from GaAs quantum wells [17] using steady state spectroscopy combined with high magnetic fields. In other reported cases, the experimental evidence (lack of a unique $\tau_D$ for example) has not substantiated SF. These include: ruby:$Cr^{3+}$ [18], GaAs laser diodes [19], ZnO nano-materials [20,21], CuCl quantum dots [22], diphenyl:pyrene [23], *R*-phycoerythrin [24], thiophene/phenylene co-oligomer [25].

ZnTe crystals, as the model II-VI direct-gap intrinsic semiconductor, have been used to attempt to observe BEC of excitons [26], and efforts have been made to resolve the lasing behaviour from highly excited ZnTe epitaxial layers with a $\tau_p$=70 ps [27]. Here we report exciton dynamics with femtosecond resolution in ZnTe crystals at 5 K studied by the femtosecond time-resolved fluorescence technique [28], showing clear observation of SF emission. A characteristic $\tau_D$, together with the associated features of quadratic power dependence, quantum noise and quantum fluctuations, quantum beating and ringing, are all observed.

The laser source used in our experiment is a commercial $\tau_p$=180 fs amplifier (Coherent RegA 9000), which delivers 5.0 μJ pulse at 100 KHz and 780 nm (1.59 eV). The pump laser at $\lambda_{ex}$=390 nm (3.18 eV), the second harmonic (SHG) of 780 nm, is focused onto the sample ZnTe crystal, at an incident angle of ~3°, the forward emission is collected and converged onto a BBO crystal, where the spontaneous emission is up-converted in a cross-linear geometry by a gating beam at 780 nm, through sum-frequency generation (SFG). The up-converted beam in the UV passes through a double-grating monochromator and reaches a solar-blind photomultiplier tube (PMT). The signal intensity from the PMT is recorded by a gated photon counting technique (Becker & Hickl PMS 400A) with respect to the time delay between pump and gating pulses. The computer controlled motorized linear translation stage (Newport IMS600) provides the minimum time delay at 8.3 fs/step. The typical response time of the system is $\Delta t$= 360 fs, shown in Fig. 1 as the $t_0$ pulse, which is measured by the SFG of pump scattering and gating beam; the spectral response bandwidth measured is ~3 nm. A fibre-coupled CCD spectrometer (Ocean Optics USB 4000) is used to simultaneously monitor the backward emission from the sample to record the steady state spectra. The pump power ($P$) is controlled with a variable neutral density filter, $P$= 10 mW means that a single pulse has $2.0 \times 10^{11}$ photons, the corresponding transient power density is 6.37 GW cm$^{-2}$; other values of $P$ have a linear relationship to this. An OD=1.0 neutral filter is used to attenuate the strong SF emission to avoid possible saturation of the detector.

Figure 1 shows the behaviour of the strong green emission from a 0.5 mm thick ZnTe crystal (110) at 5 K excited with a femtosecond pulse at $\lambda_{ex}$=390 nm (3.18 eV). At very low $P$ the bound exciton emission lines can be recognized as two main peaks at 522 nm (2.375 eV, I1) and 525 nm (2.360 eV, I2) respectively, and a weak shoulder at 532 nm (2.331 eV, I3) (Fig. 1(a)&(b)) [26]. As $P$ increases, exciting some areas of the sample, the I1 peak dominates the emission spectrum and slightly red-shifts to 522.5 nm (2.373 eV) from 521.5 nm (2.378 eV), due to a band-gap renormalization effect over the dense exciton population, $N$ (Fig. 1(a)), the I1 linewidth broadens to 6.0 nm (27 meV) at 22 mW from 2.8 nm (13 meV) at low $P$: the $P$ dependence of its peak intensity, plotted in Fig. 1(c), shows a linear growth; the corresponding ultrafast emission dynamics in Fig. 1(d) shows ordinary build-up followed by accelerated exciton recombination on $P$. This is typical of an electron-hole plasma (EHP) [15]. Whereas exciting other areas of the sample, we observe the onset of "lasing" for the I1 line as $P>P_0$, as observed before [27], now the I1 peak has greatly increased intensity and undergoes a large red-shift to 524.5 nm (2.364 eV) at $P$=18 mW and its linewidth narrows to 3.2 nm (14 meV) (Fig. 1(b)). Fitting the $P$ dependence of peak intensities in Fig. 1(c) gives an excellent quadratic relation as $P>P_0$. The $P$ dependent ultrafast dynamics of lasing at ~524 nm (2.365 eV) are depicted in Fig. 1(e): the period from pump pulse ($t_0$) to build-up time ($t_{ref}$) represents the initial processes of carrier scattering for the release of excess energy, and following exciton formation, resulting in an *incoherent hot exciton* population $N$. At low $P$ the curves are similar to those shown in Fig. 1(d); but, surprisingly, as $P>P_0$, new high intensity peaks emerge riding on top of the spontaneous emission with a clear time delay, $\tau_D$, with respect to $t_{ref}$. The time delay, $\tau_D$, decreases gradually from 40 ps at $P$=6.0 mW to 10 ps at 18 mW, this is unique characteristic of SF [5-10]. Given a logarithmic intensity scale in Fig. 1(e) one can easily see the relatively noisy signals under moderate $P$ compared to the curves at low $P$: this feature is typical of the macroscopic appearance of quantum fluctuations, again characteristic of SF, *i.e.* quantum noise [3,5], resulting from the multi-shot measurements [5,9]. Furthermore, at higher $P$ (>10 mW), regular interference fringes in the vicinity of the maximum of each curve are observed characteristic of quantum beating of SF among multiple SF modes [7-9].

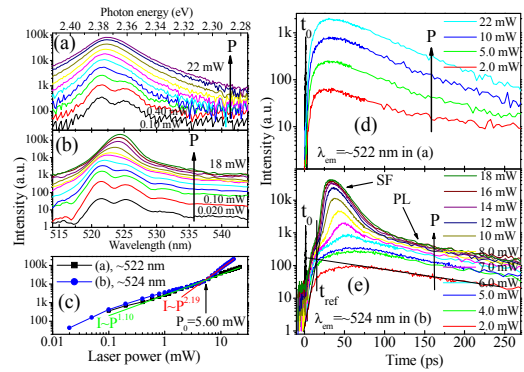

FIG. 1 (color online). Spontaneous photoluminescence (PL) from electron-hole plasma (EHP) *versus* SF in a 0.5 mm ZnTe single crystal at 5 K excited by a femtosecond laser at $\lambda_{ex}$=390 nm (3.18 eV). Power ($P$) dependent emission spectra from non-lasing and lasing areas are shown in (a) & (b) respectively. (c) $P$ dependence of peak intensities in (a) and (b). (d) $P$ dependent ultrafast emission dynamics of EHP. (e) $P$ dependent SF emission process. $t_{ref}$ is defined by the cross-point of build-up and decay as illustrated on 2.0 mW curve, $t_0$ pulse is the excitation pulse.

These observations are very reproducible; Fig. 2 shows the results from another ZnTe crystal with a thickness of 0.2 mm. As $P$ is much higher, the SF peak sweeps through I2 and gradually red-shifts to 528 nm (2.348 eV) at $P$ =50 mW, leaving I1 behind as a shoulder (Fig. 2(a)). The $P$ dependence in Fig. 2(b) also indicates a clear quadratic growth in comparison to I1 as $P>P_0$. Importantly, in Fig. 2(c), besides the reduction of $\tau_D$ with $P$ and the presence of quantum noise and fringes, each SF trace has two maxima as $P$>30 mW: this is the characteristic "ringing" of SF [7-9]. Moreover, the fringes are sufficiently clear that a beat frequency of ~1.8 THz can be recognized from the corresponding fast Fourier transform (FFT) spectra in Fig. 2(d), which is the difference of two SF frequencies [7-9].

SF not only takes place on I1 but also can be observed on I2: the data recorded are shown in Fig. 3, in which the general features of SF are all observed as in Figs. 1&2. In this case the SF peak has a huge red-shift (Fig. 3(a)), through I3, to 540 nm (2.296 eV) at $P$=50 mW (from 526 nm (2.357 eV) at 4.0 mW). The $P$ dependence of the emitted intensity shown in Fig. 3(b) also has an excellent quadratic relationship with respect to I1 as $P>P_0$. The deviation as $P>10$ mW is attributed to the gain saturation effect and possible damage by the high power density of the pump laser. A most interesting feature is the $P$-dependent quantum beating fringes in Fig. 3(c) and their FFT spectra in Fig. 3(d). At moderate $P$, the large period fringes reflect lower beat frequencies: 0.38 THz at $P$=4.0 mW and 0.78 THz at 6.0 mW and 8.0 mW. As $P$ increases, the fringes become dense and beating shifts to 4.15 THz at 10 mW and 12 mW; when $P>12$ mW; the beat frequency shifts to ~5.5 THz and many beats appear. This is because the multiple SF frequencies are activated as the SF peak shifts away from I2, as shown in Fig. 3(a).

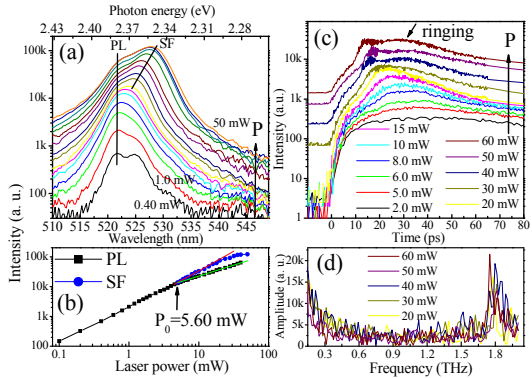

FIG. 2 (color online). SF emission on I1 line from a 0.2 mm ZnTe single crystal at 5 K excited by a femtosecond laser at $\lambda_{ex}$= 390 nm (3.18 eV). (a) $P$ dependent emission spectra, the guide lines showing PL I1 peak and shift of SF peak respectively. (b) Comparison of $P$ dependence of peak intensities in (a). (c) $P$ dependent SF emission dynamics, some curves are up-shifted for clarity. (d) The corresponding FFT spectra of the curves in (c).

SF emission weakens rapidly as temperature increases, and disappears completely at ~45 K at the maximum $P$, these are similar findings to those in [10].

Given the above observations, SF from the highly excited buck ZnTe single crystals at 5 K can be confirmed unambiguously with the evidence of a clear characteristic induction time, $\tau_D$, quantum fluctuations and noise signals, quantum beating and beats, and ringing, in addition to emission line narrowing and quadratic $P$ dependence of $I_{SF}$ [2-10]. To the best of our knowledge, this is the second fully evidenced report of SF in the solid state, after that in KCl:$O_2^-$ [10,11]; and the first unambiguous report of SF in an intrinsic semiconductor with the key evidence in the time domain. Our observations further support those by Jho *et al* made in the frequency domain [17].

As SF arises from $N_c$, a small subset of $N$, one can note that even at higher $P$, the whole emission spectra does not collapse significantly into a single peak, but is a mixture of SF emission from $N_c$ and spontaneous emission from the incoherent excitons ($N$-$N_c$), the ultrafast emission dynamics above also show the clear difference between $N_c$ and ($N$-$N_c$). These observations and the quantum beating and ringing also noted, indicate that 'lasing' here is clearly not ASE but SF; the $\tau_D$ process also excludes SR.

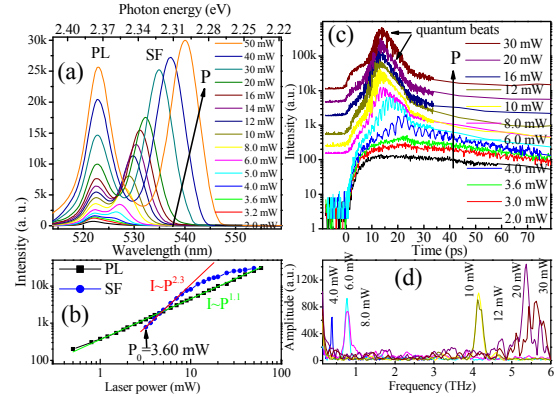

FIG. 3 (color online). SF emission on I2 line from a 0.2 mm ZnTe single crystal at 5 K excited by a femtosecond laser at $\lambda_{ex}$= 390 nm (3.18 eV). (a) $P$ dependent emission spectra. (b) Comparison of $P$ dependence of the peak intensities in (a). (c) $P$ dependent SF emission dynamics, some curves are up-shifted for clarity. (d) The corresponding FFT spectra of the curves in (c).

From the results above we can estimate the proportion of $N_c$ in $N$. The excitation laser has a spot diameter of 100 μm and a penetration depth $l$≈0.5 μm, giving an excitation volume $V$≈4.0×10$^3$ μm$^3$. Fitting the curves at low $P$ in Fig. 1(e) gives a typical radiative lifetime $\tau_{rad}$=120 ps. With $\Phi$≈10%, the estimated quantum yield of green emission from ZnTe at 5 K, then the spontaneous lifetime of excitons is $\tau_{SP}$≈ 1,200 ps. Using $\lambda$= 524 nm, $\tau_R$=(8$\pi$/3$N_c\lambda^2 l$)$\tau_{SP}$ and $\tau_D$= $\tau_R[ln(2\pi SlN_c)^{1/2}]^2/4$, we find (i) when $\tau_D$=40 ps at $P$=6.0 mW, $N_c$=5.0×10$^4$ μm$^{-3}$, $\tau_R$=1.5 ps, and ($\tau_R\tau_D$)$^{1/2}$=7.7 ps; (ii) for $\tau_D$=10 ps at $P$=18 mW, $N_c$= 2.3×10$^5$ μm$^{-3}$, $\tau_R$=0.32 ps, and ($\tau_R\tau_D$)$^{1/2}$=1.8 ps. Given the estimated $T_2^*$≈10 ps for excitons in ZnTe at 5 K [28,29], then the criterion ($\tau_R\tau_D$)$^{1/2}$ <$T_2^*$ for SF is satisfied. Using $\Phi$≈10% again, and taking the photon number of a single pump pulse, $N$ ≈1.2×10$^{10}$ /V ≈3×10$^6$ μm$^{-3}$ at 6.0 mW or $N$≈9×10$^6$ μm$^{-3}$ at 18 mW (similar order of magnitude to the Mott density, 10$^6$ μm$^{-3}$), yielding ratios of $N_c/N$ as approximately 1.7% and 2.6% respectively. More detailed analysis with a full theoretical model is underway and is not included here. Furthermore, the ultrashort $\tau_R$ value indicates the linewidth of SF emission is a result of both the lifetime broadening and the multi-shot measurement; the SF pulse shape recorded thus reflects the interplay of $\tau_D$ and $\tau_R$.

Of particular and greater significance, the occurrence of SF signifies the spontaneous macroscopic quantum coherence of an ensemble of excitons, $N_c$, formed within the clear induction time $\tau_D$, after a femtosecond excitation pulse creates thermalized incoherent excitons $N$. We think that the mechanism for the spontaneous coherence is the formation of a BEC of excitons on this ultrashort timescale in the intrinsic ZnTe single crystal at 5 K [12-16, 30].

Under our experimental conditions, within the small excitation volume ($V \approx 4.0 \times 10^3$ μm$^3$) in some regions of the crystal bulk, a high density ensemble of bound excitons can form having sufficiently long $T_2^*$ that spontaneous symmetry breaking within a time $\tau_D$ forms a spontaneous coherence in a single quantum state I1 or I2 *i.e.* a BEC of excitons forms. The whole coherent system then being triggered by a random spontaneous emission event, yielding a burst off observable SF pulses when the coherent exciton density $N_c$ exceeds a certain threshold as $P > P_0$. This case is clearly different from the so-called "driven-BEC" of exciton-polaritons in microcavities [14-16].

In summary we have observed clear evidence of superfluorescence in bulk ZnTe crystals between 5 K and 45 K under femtosecond laser excitation. A clear characteristic induction time, $\tau_D$, together with quantum noise, quantum beating and ringing support this unambiguous observation of SF emission. We put forward that this very possibly is also evidence that a BEC of excitons has formed, which then spontaneously decays to yield the SF burst and so it is the process of spontaneous Bose condensation that governs the formation of the coherent ensemble of excitons in the ZnTe crystals on an ultrafast timescale. Our case is different from other cases of "driven-BEC". The decoherence factor $T_2^*$, which is related to the degree of perfection of the crystal structure, must be considered in achieving BEC of excitons. The careful use of femtosecond time-resolution techniques to identify the SF process on an ultrafast timescale should enable future experimental studies in other semiconductor materials.

We thank Profs. R. A. Abram and J. M. Chamberlain for fruitful discussions and critical review.

Figure captions:

FIG. 1 (color online). Spontaneous photoluminescence (PL) from electron-hole plasma (EHP) *versus* SF in a 0.5 mm ZnTe single crystal at 5 K excited by a femtosecond laser at $\lambda_{ex}$=390 nm (3.18 eV). Power (*P*) dependent emission spectra from non-lasing and lasing areas are shown in (a) & (b) respectively. (c) *P* dependence of peak intensities in (a) and (b). (d) *P* dependent ultrafast emission dynamics of EHP. (e) *P* dependent SF emission process. $t_{ref}$ is defined by the cross-point of build-up and decay as illustrated on 2.0 mW curve, $t_0$ pulse is the excitation pulse.

FIG. 2 (color online). SF emission on I1 line from a 0.2 mm ZnTe single crystal at 5 K excited by a femtosecond laser at $\lambda_{ex}$=390 nm (3.18 eV). (a) *P* dependent emission spectra, the guide lines showing PL I1 peak and shift of SF peak respectively. (b) Comparison of *P* dependence of peak intensities in (a). (c) *P* dependent SF emission dynamics, some curves are up-shifted for clarity. (d) The corresponding FFT spectra of the curves in (c).

FIG. 3 (color online). SF emission on I2 line from a 0.2 mm ZnTe single crystal at 5 K excited by a femtosecond laser at $\lambda_{ex}$=390 nm (3.18 eV). (a) *P* dependent emission spectra. (b) Comparison of *P* dependence of the peak intensities in (a). (c) *P* dependent SF emission dynamics, some curves are up-shifted for clarity. (d) The corresponding FFT spectra of the curves in (c).

Fig. 1.

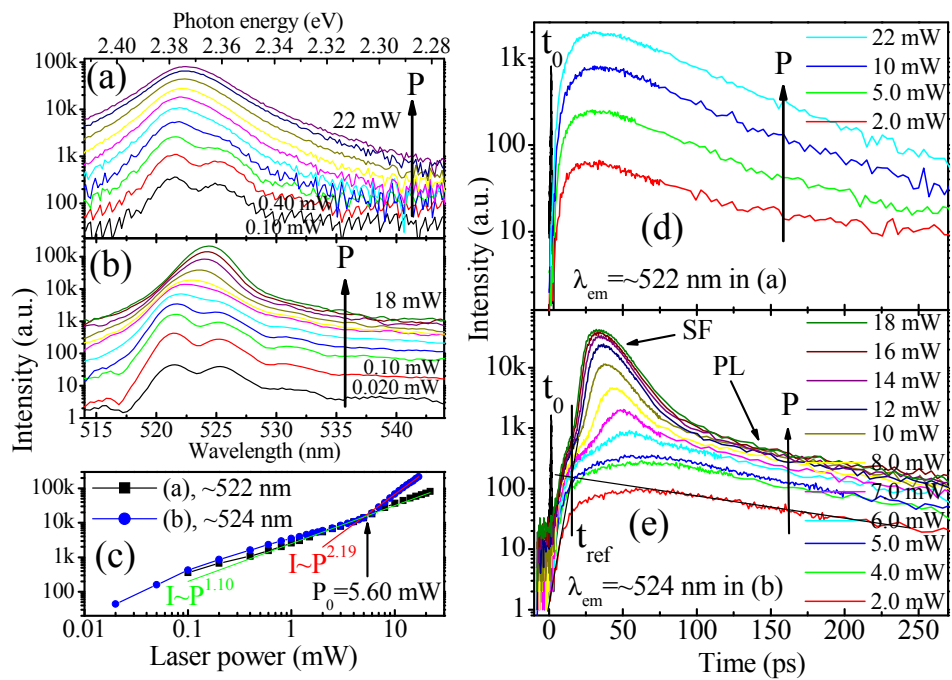

Fig. 2.

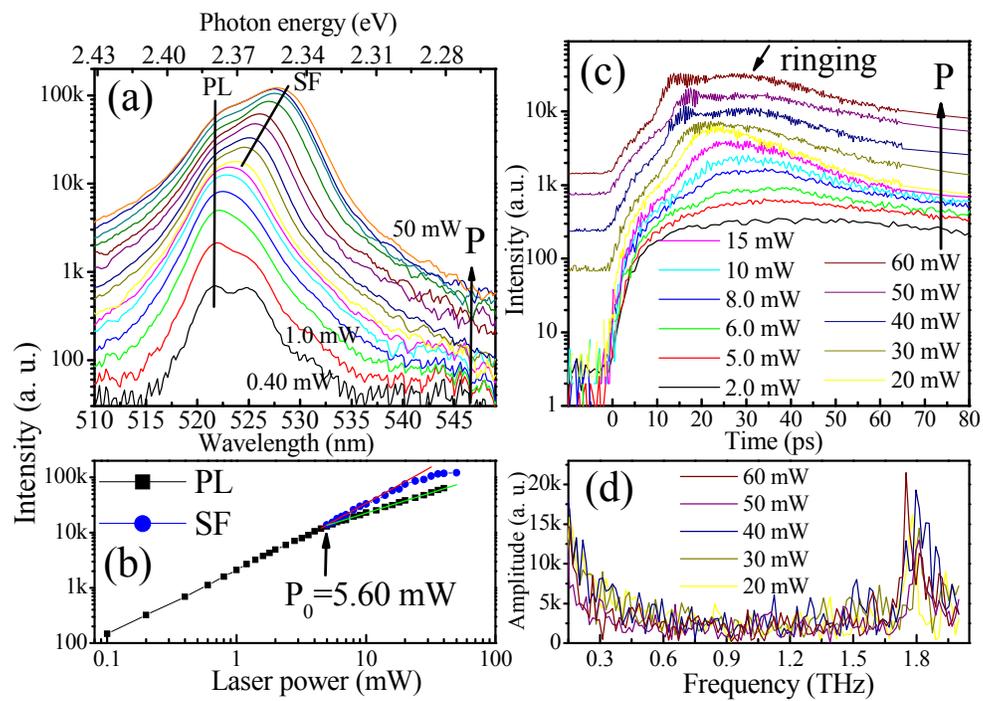

Fig. 3.

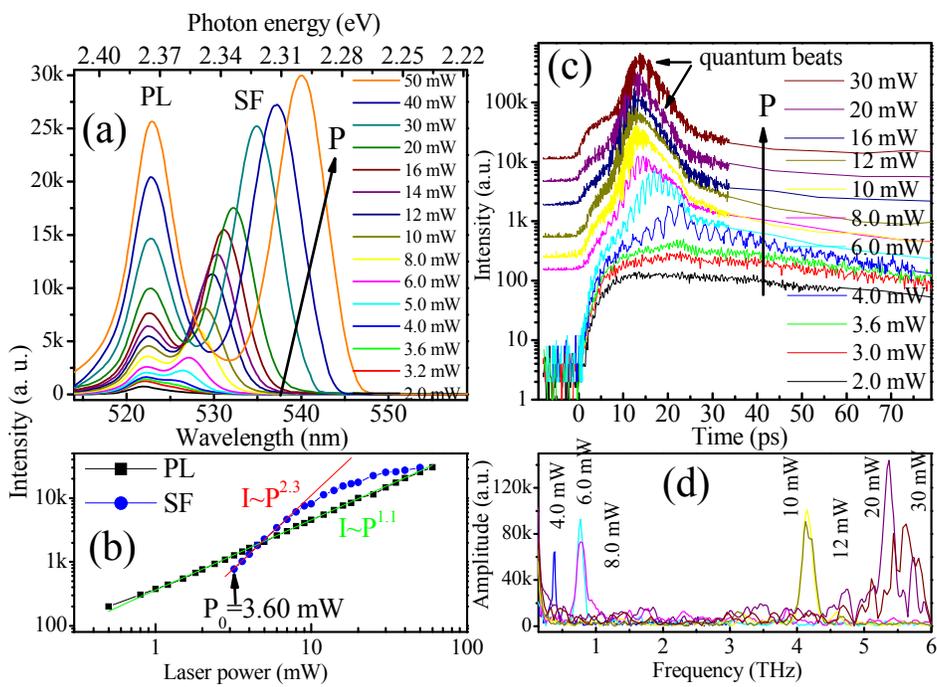

8